\documentclass[a4paper,11pt]{article}
\pdfoutput=1
\usepackage{jcappub} 
\usepackage[T1]{fontenc} 

\usepackage{mathtools}
\usepackage{bm}
\usepackage{comment}

\title{Remarks on overestimating the effects of inhomogeneities on the Hubble constant}

\author[a,1]{Taishi Miura,\note{Corresponding author.}}
\author[a,b]{Takahiro Tanaka}
\affiliation[a]{Department of Physics, Kyoto University, Kyoto 606-8502, Japan}
\affiliation[b]{Center for Gravitational Physics and Quantum Information, Yukawa Institute for Theoretical Physics, Kyoto University, Kyoto 606-8502, Japan}

\emailAdd{miura@tap.scphys.kyoto-u.ac.jp}
\emailAdd{t.tanaka@tap.scphys.kyoto-u.ac.jp}

\abstract{
The Hubble constant is one of the most important parameters in cosmology.
Discrepancies in values of the Hubble constant estimated from various measurements, the so-called Hubble tension, are a serious problem.
In this paper, we study the effects of small-scale inhomogeneities of structure formation on the measurement of the Hubble constant using the luminosity distance-redshift relation.
By adopting the adhesion model in Newtonian cosmology as the model of structure formation, we investigate whether or not the effects of inhomogeneities can be sufficiently large to affect the current observations of the Hubble constant.
We show that inappropriate treatment of the effects of inhomogeneities can cause a large deviation of the measured value of the Hubble constant from the background value, whose magnitude is comparable with the Hubble tension. 
Our main message is the importance of adopting an appropriate model of structure formation to investigate the effects of inhomogeneities.
We also add discussion on the spatial averaging approach used to estimate the measured Hubble constant in the inhomogeneous universe.
}

\begin{document}

\maketitle
\flushbottom

\section{Introduction}

The concordance \( \Lambda \)CDM (Cold Dark Matter) model has explained various cosmological observations.
However, some inconsistencies in the \( \Lambda \)CDM model have emerged in the precision cosmology era.
One of the most significant inconsistencies is about the Hubble constant \(H_0\).
The local measurement from the distance-redshift relation using Cepheid variable stars and Type Ia supernovae gives \(H_0 = 73.04 \pm 1.04 \, \mathrm{km} \, \mathrm{s}^{-1} \mathrm{Mpc}^{-1} \)~\cite{Riess:2021jrx}.
This measurement does not require any particular assumptions on cosmological models.
The Planck measurement of the cosmic microwave background (CMB), in which the \( \Lambda \)CDM model is assumed, estimates \(H_0 = 67.36 \pm 0.54 \, \mathrm{km} \, \mathrm{s}^{-1} \mathrm{Mpc}^{-1} \)~\cite{Planck:2018vyg}.
The magnitude of this discrepancy is about \(10 \% \) and \(5 \sigma \).
Other measurements of the Hubble constant~\cite{Wong:2019kwg, Freedman:2019jwv, eBOSS:2020yzd, DES:2017txv, LIGOScientific:2017adf,Perivolaropoulos:2021jda,Lenart:2022nip,Bargiacchi:2023jse} have been done.
While their estimates are still less precise than these two, 
they suggest the existence of an inconsistency between observations of the early universe and those of the late universe~\cite{Verde:2019ivm,DiValentino:2021izs}.
The results of the various cosmological observations are examined in detail in relation to this tension among the values of the observed Hubble constants called as the Hubble tension~\cite{Dainotti:2021pqg,Dainotti:2022bzg,Dainotti:2023ebr,Bargiacchi:2023rfd}.
If the observational results are correct, the cosmological model may need to be extended from the \( \Lambda \)CDM model.
A lot of ideas of extending the \( \Lambda \)CDM model to solve the Hubble tension have been proposed and discussed~\cite{DiValentino:2021izs,Mortsell:2018mfj,Knox:2019rjx,Arendse:2019hev,Jedamzik:2020zmd, Lin:2021sfs,Schoneberg:2021qvd,Cai:2021wgv,Krishnan:2021dyb}.

Although whether the cosmological principle is valid or not is also questioned~\cite{Aluri:2022hzs}, the universe is usually thought to be approximated by a homogeneous and isotropic spacetime model on large scales.
Adding small linear perturbations on the homogeneous and isotopic universe would be enough to describe the real universe on large scales.
On small scales, however, the magnitude of perturbations becomes large because of the cosmological structure formation, and the perturbative treatment breaks down.
It has been argued that the global deviation from the Friedmann-Lemaître-Robertson-Walker (FLRW) model caused by the effects of inhomogeneities, the so-called cosmological backreaction, may change the measured values of the cosmological parameters significantly~\cite{Rasanen:2006zw, Wiltshire:2007jk, Buchert:2011sx, Roukema:2013cya, Buchert:2015iva, Santos:2016sog, Roukema:2017doi}, mainly in the context of proposing a possible explanation of the accelerated expansion of the universe without dark energy.
While it has also been claimed that the effects of inhomogeneities of the universe are not sufficient to explain the accelerated expansion of the universe~\cite{Ishibashi:2005sj, Green:2010qy, Green:2011wc, Ben-Dayan:2013nkf, Adamek:2014qja},
some works claim that the non-linear effects may change the global expansion rate and solve the Hubble tension~\cite{Tomita:2017eem, Tomita:2017bpz, Bolejko:2017fos, Heinesen:2020sre}.
By contrast, some works claim that the effect of the backreaction does not change the measured value of the Hubble constant significantly~\cite{Adamek:2018rru, Tian:2020qnm}.
Many other studies have investigated the effects of inhomogeneities on the cosmological observables which are essential for measuring the Hubble constant and the accelerated expansion~\cite{Futamase:1989hba, Kaiser:2015iia, Fleury:2016fda, Breton:2020puw,Rasanen:2009uw,Rasanen:2008be,Koksbang:2020zej,Koksbang:2019glb,Koksbang:2019wfg,Koksbang:2019cen,Koksbang:2020qry,Koksbang:2021zyi,Umeh:2022kqs,Umeh:2022prn,Umeh:2022hab,Yu:2022wvg,Fanizza:2019pfp,Schiavone:2023olz,Ben-Dayan:2014swa,Fanizza:2021tuh}.
It is true that it is interesting if the backreaction effect could solve the cosmological problems without requiring any extensions from the \( \Lambda \)CDM model.
In this paper we critically reexamine whether or not the measured values of the Hubble constant can be affected by the inhomogeneities largely enough to explain the Hubble tension.

The effects of inhomogeneities are often discussed based on averaging a certain quantity over a constant time hypersurface.
It is also discussed how the cosmological gauges adopted for this averaging can affect the results~\cite{Adamek:2017mzb, Buchert:2018yhd}.
However, actual cosmological measurements utilize light rays that propagate over cosmological distances.
Thus, the averaged quantities over a constant time hypersurface are not corresponding to observables of cosmological measurements,
and the errors caused by a naive averaging can be uncontrollable in the estimate of the Hubble constant.
In addition, when studying the effects of inhomogeneities, the structure of the spacetime is sometimes replaced by a simple model~\cite{Roukema:2017doi, Bolejko:2017wfy}.
These treatments of the structure of the spacetime may lead to unphysical results.

Hence, in this work we examine the effects of inhomogeneities of the universe on the local measurement of the Hubble constant using the luminosity distance-redshift relation as is done in the actual measurement in~\cite{Riess:2021jrx}.
To properly study the Hubble tension, we also need to compare the results with the estimations of the Hubble constant using other observations such as the CMB observation.
In this paper, however, we focus on whether the Hubble constant measured from the luminosity distance-redshift relation deviates significantly from the background value, since our final conclusion to this question is negative. 
We calculate the luminosity distance-redshift relation considering the null congruence
and evaluate the effects of inhomogeneities on the Hubble constant estimation.
For comparison, we change the treatment of the non-linear structure formation 
and estimate how the estimated values of the Hubble constant varies.
From this analysis, we demonstrate the importance of careful treatment of the modeling of the gravitationally 
collapsed regions in examining the effects of inhomogeneities.
Although one may think that solving exact non-linear evolution of inhomogeneities within general relativity is important for discussing the effects of inhomogeneities, we consider the non-linear evolution of the small-scale inhomogeneities within the cosmological Newtonian approximation. 
The adhesion model and the freezing model are adopted in this work to model the cosmological structure which would be well described within the cosmological Newtonian approximation. 
We adopt this Newtonian approximation because the difference between the different treatments on the measured values of the Hubble constant can be clearly seen within this approximation.
The motivation of this works is to emphasize the importance of how to treat gravitationally collapsed regions even within the Newtonian approximation, rather than showing that the Newtonian approximation is sufficient to estimate the effect of inhomogeneities on the measured Hubble constant.
If the structure formation were modeled in similar ways even in general relativity, 
the results should be close to those obtained in this work, because the difference depending on the treatment appears at the level of the Newtonian approximation.
Finally, we also give a brief comment on an averaging approach based on our model, demonstrating the errors possibly contained in this approach.

The paper is organized as follows.
In Section~\ref{sec:h0estimation}, we explain our model of the structure formation, 
derive the equations necessary to evaluate the Hubble constant, 
and show the results of the effects of small-scale inhomogeneities.
In Section~\ref{sec:freezing}, we consider how the results change if an inappropriate model of the structure formation is adopted.
We discuss the averaging approach in Section~\ref{sec:averaging}.
Finally, we conclude the paper in Section~\ref{sec:conclusion}.
The code used to obtain the results in this work is available on GitHub.\footnote{\url{https://github.com/tmiura79/IHC}}

\section{Hubble constant estimation \\by the luminosity distance-redshift relation}\label{sec:h0estimation}

In this section, we study the effects of small-scale inhomogeneities on the measurement of the Hubble constant and show that they are negligible in Newtonian cosmology.
Here, we introduce a simple model of non-linear structure formation in the universe.
In this model, we calculate the Hubble constant determined by the local observations using the luminosity distance-redshift relation.

\subsection{Zel'dovich approximation}\label{sec:metric}
We use the Newtonian gauge and consider only the scalar-type perturbations. 
The vector and tensor-type perturbations are neglected.
Then, the infinitesimal line element is given by 
\begin{align} \label{eq:newtoniangauge}
    ds^2 = -(1+2\phi(t,x^k)) dt^2 + a^2(t) (1-2\psi(t,x^k)) \delta_{ij} dx^i dx^j ,
\end{align}
where \(t\) is the cosmic time, \(\{x^i\}\) are the comoving spatial coordinates, and \(a(t)\) is the scale factor. 
Here, we assume that a flat FLRW universe and the energy content of the universe consists only of the dust fluid 
component. 
Thus, \(a(t) \propto t^{2/3}\) and the energy-momentum tensor is
\begin{align}\label{eq:dustEMtensor}
    T_{\mu \nu} = \rho u^{\mu} u^{\nu},
\end{align}
where \( \rho \) is the energy density of the dust fluid in the rest frame, and \( u^{\mu} \) is the 4-velocity of the dust fluid.
In the Newtonian gauge, the 4-velocity \(u^{\nu}\) is written as
\begin{align}
    u^0 = \frac{1}{\sqrt{1 + 2 \phi - a^2(1 - 2 \psi) v^2}}, \label{eq:4velocity0}\\
    u^i = \frac{v^i}{\sqrt{1 + 2 \phi - a^2(1 - 2 \psi) v^2}}, \label{eq:4velocityi}
\end{align}
where \( v^i \equiv \displaystyle{dx^i(t)}/{dt} \) is the velocity of the fluid. 
\(x^i(t)\) denotes the trajectory of the fluid element parameterized by the cosmic time \(t\), and \(\displaystyle{d}/{dt}\) is the time derivative with respect to the cosmic time \(t\) along the world line of each fluid element.

Here, we regard \( \phi \) and \( \psi \) as small perturbations, and assume that the dust motion is nonrelativistic:
\begin{align}
    av \ll 1\,.
\end{align}
Furthermore, the time variation of the fields is in general assumed to be slow
\begin{align}
    {\left(\frac{\partial}{\partial t}\right)}_{x^i} =\mathcal{O} (v) {\left( \frac{\partial}{\partial x^i} \right)}_{t}.
\end{align}
We also restrict our attention to the case in which the length scale of perturbation $a L$ is sufficiently short compared with the Hubble length scale:
\begin{align}
    aHL\ll 1,
\end{align}
where \(H\) is the Hubble parameter.
We assume that the small quantities satisfy the following order of magnitude relations,
\begin{align}
    \phi, \psi, {(av)}^2, {\left( aH L\right)}^2 =\mathcal{O}(\epsilon^2),
\end{align}
with \(\epsilon \) being a small parameter. 
The approximate equality ${\left( aHL\right)}^2\sim \phi,\psi$ follows from the condition that the gravitational collapse proceeds in the time scale of the age of the universe. 
We assume that the magnitude of the density perturbation \( \delta\rho \equiv \rho - \bar{\rho}(t)\) can be \(\mathcal{O}(1)\), where \( \bar{\rho}(t)\) is the energy density of the background universe.

Calculating the time-time and space-space components of the Einstein equation in the Newtonian gauge, and using the background equations, 
we obtain
\begin{align}
    \Delta_{\bm{x}} \psi &= 4 \pi G a^2 \delta \rho + L^{-2}\mathcal{O}(\epsilon^4), \label{eq:NGeq}\\
     \phi &= \psi + \mathcal{O}(\epsilon^4), \label{eq:NGeq2}
\end{align}
where \( \Delta_{\bm{x}} \coloneqq \partial^2/\partial x^2 + \partial^2/\partial y^2 + \partial^2/\partial z^2\) is the Laplacian with $t$ kept constant.

Calculating the conservation law of the energy-momentum tensor using Eq.~\eqref{eq:dustEMtensor}, we get
\begin{align}
   { \left( \frac{\partial \rho}{\partial t} \right)}_{\bm{x}} &+ 3 H \rho + \nabla_{\bm{x}} \cdot (\rho \bm{v}) = L^{-1}\mathcal{O}(\epsilon^3), \label{eq:NGfliuidcontinuityeq}\\
    {\left( \frac{\partial \bm{v}}{\partial t} \right)}_{\bm{x}} &+ 2H\bm{v} + (\bm{v} \cdot \nabla_{\bm{x}}) \bm{v} = - \frac{1}{a^2} \nabla_{\bm{x}} \phi + L^{-1}\mathcal{O}(\epsilon^4), \label{eq:NGfluidEulereq}
\end{align}
where \(\nabla_{\bm{x}} \coloneqq \left({\partial}/{\partial x}, {\partial}/{\partial y}, {\partial}/{\partial z}\right)\) is the nabla operator with respect to \(x^i\), holding \(t\) constant.
The time-space components of the Einstein equation are not necessary because they can be derived from Eqs.~\eqref{eq:NGeq}-\eqref{eq:NGfluidEulereq}.
Eqs.~\eqref{eq:NGeq}-\eqref{eq:NGfluidEulereq} take the same forms as those of Newtonian cosmology.
Therefore, it is obvious 
that the backreaction effects of the structure formation on the metric are 
$\mathcal{O}(\varepsilon^4)$ or higher.

To demonstrate that the Hubble constant determined from measurable quantities cannot be amplified as long as $\varepsilon$ remains small,  
we use the Zel'dovich approximation~\cite{Zeldovich:1969sb}. 
In this approximation, the comoving coordinates \(\bm{x}\) and the Lagrangian coordinates \(\bm{q}\) are related with each other by  
\begin{align}\label{eq:zeldovich}
\bm{x}(t,\bm{q}) = \bm{q} - b(t) \,\nabla_{\bm{q}} \Phi(\bm{q}) ,
\end{align}
where \(b(t)\) is the linear growth factor and \(\Phi(\bm{q})\) is the velocity potential. 

In the dust-dominant universe, \(b(t) \propto a(t)\).
In addition, we consider the plane-symmetric case, where the velocity potential depends only on one component of \(\bm{q}\). (We choose it as \(x\)-direction.)
In this case, the Zel'dovich approximation is an exact solution until caustics, where neighboring fluid elements mutually intersect, are formed.
From Eq.~\eqref{eq:zeldovich}, the \(x\)-component of the velocity field \(v\) is 
\begin{align}\label{eq:velocity}
v(t,x(t,q)) = {\left(\frac{\partial x(t,q)}{\partial t}\right)}_q = - \dot{b}(t)\, \frac{d \Phi(q)}{d q} ,
\end{align} 
where \(\dot{~} \equiv {d}/{d t}\).
From the mass conservation, we find that the density is given by 
\begin{align}\label{eq:density-newton}
    \frac{\rho(t,x(t,q))}{\bar{\rho}(t)}  
    \equiv 1 + \delta(t,x(t,q)) 
    = {\left[\mathrm{det}\left(\frac{\partial \bm{x}}{\partial \bm{q}}\right)\right]}^{-1}
    =\frac{1}{1 - b(t)\, \displaystyle\frac{d^2 \Phi}{dq^2}}\,.
\end{align}
We can see that Eqs.~\eqref{eq:velocity} and~\eqref{eq:density-newton} satisfy Eq.~\eqref{eq:NGfliuidcontinuityeq}.
Substituting Eq.~\eqref{eq:velocity} into the Euler equation \eqref{eq:NGfluidEulereq}, 
we obtain 
\begin{align}\label{eq:ZAonephi}
\phi(t,x(t,q)) = \frac{3}{2} \dot{a}^2(t) b(t) \left( \Phi(q) - \frac{b(t)}{2} {\left( \frac{d \Phi}{dq} \right)}^2\right)\,,
\end{align}
which consistently solves the Poisson equation \eqref{eq:NGeq} with the substitution of the density field \eqref{eq:density-newton}.

\subsection{Adhesion model}\label{sec:adhe}
In the Zel'dovich approximation, motions of fluid elements are  determined only by the local initial conditions, 
and the approximation does not make sense 
when the gravitational attraction starts to form caustics because the map from 
$\bm{x}$ to $\bm{q}$ becomes multi-valued.
Here, we use the adhesion model \cite{Gurbatov:1989az,Schandarin:1989sr}, in which infinitesimal  artificial viscosity is added to the Zel'dovich approximation to 
mimic the relaxation on small scales due to self-gravity.

In the Zel'dovich approximation, the velocity \(\bm{v}\) is expressed as Eq.~\eqref{eq:velocity}.
It means that when we adopt the growth rate \(b(t)\) as the time coordinate, 
the fluid elements behave as if they do not experience any forces.
Also, it tells that the new quantity defined by \(\bm{U}(t,\bm{x}(t,\bm{q})) \equiv \bm{v}(t,\bm{x}(t,\bm{q}))/\dot{b}(t)\) depends only on the Lagrange coordinates \(\bm{q}\).
Using  \(b(t)\) and \(\bm{q}\) as the independent spacetime coordinate variables, we get
\begin{align}
    \label{eq:Eulereqzeldo1}
    {\left(\frac{\partial \bm{U}(b,\bm{q})}{\partial b} \right)}_{\bm{q}} = 0. 
\end{align}
Choosing \(b\) and \(\bm{x}\) as the independent spacetime coordinate variables, 
Eq.~\eqref{eq:Eulereqzeldo1} is rewritten as 
\begin{align}\label{eq:Eulereqzeldo2}
    {\left(\frac{ \partial \bm{U}}{\partial b}\right)}_{\bm{x}} +  (\bm{U} \cdot \nabla_{\bm{x}} ) \bm{U} = 0.
\end{align}
As mentioned above, the Zel'dovich approximation does not make sense after the formation of caustics. 
To avoid this, in adhesion model an artificial viscosity is added to Eq.~\eqref{eq:Eulereqzeldo2}
as
\begin{align}
    {\left(\frac{\partial \bm{U}}{\partial b} \right)}_{\bm{x}} + (\bm{U} \cdot \nabla_{\bm{x}})\bm{U} = \nu \nabla_{\bm{x}}^2 \bm{U}\,,
\end{align}
where \(\nu \) is a constant kinematic viscosity parameter.
This equation has an exact solution
\begin{align} \label{eq:adhesolution}
    \bm{U}(\bm{x},b) = \frac{\displaystyle\int \frac{\bm{x}-\bm{p}}{b} \mathrm{\exp}\{ {(2\nu)}^{-1} G(\bm{x}, \bm{p}, b)\} d^3 p}{\displaystyle\int  \mathrm{\exp}\{{(2\nu)}^{-1} G(\bm{x}, \bm{p}, b)\} d^3p}\,,
\end{align}
where
\begin{align}\label{eq:Gofadhesion}
    G(\bm{x},\bm{p},b) = \Phi(\bm{p}) - \frac{{(\bm{x} - \bm{p})}^2}{2b}\,.
\end{align} 
Now, we take the limit of \( \nu \to 0 \).
This limit is different from the case setting $\nu=0$ from the beginning. 
In this limit the value of the integral in Eq.~\eqref{eq:adhesolution} 
evaluated by the stationary phase approximation, which 
comes from the absolute maximum of \(G(\bm{x}, \bm{p}, b)\), becomes exact. 
At \(\bm{p} = \bm{p}^*\), \(G(\bm{x}, \bm{p}, b)\) takes the absolute maximum. 
Thus, the necessary condition,
\begin{align}\label{eq:maxcondition}
     {\left( \frac{\partial G(\bm{x},\bm{p},b)}{\partial \bm{p}} \right)}_{\bm{x},b}\,\biggr|_{\bm{p}=\bm{p}^*(\bm{x},b)} = 0\,,
\end{align}
is satisfied. From Eq.~\eqref{eq:maxcondition}, we find
\begin{align}\label{eq:adhesionxone}
    \bm{x} = \bm{p}^*(\bm{x},b) - b \, \nabla_{\bm{p}}\Phi |_{\bm{p} = \bm{p}^*(\bm{x} ,b)}\,.
\end{align}
For a space-time point specified by \(b\) and \(\bm{x}\) that gives only one \(\bm{p}^*\), Eq.~\eqref{eq:adhesolution} is
\begin{align}\label{eq:adhesionUone}
    \bm{U}(\bm{x},b) = \frac{\bm{x} - \bm{p}^{*}}{b} = - \nabla_{\bm{p}} \Phi(\bm{p^*})\,,
\end{align}
which is identical to the expression for the case setting $\nu=0$ from the beginning, namely the Zel'dovich approximation, by regarding \(\bm{p}^*\) as the Lagrange coordinate \(\bm{q}\).

\begin{figure}[tbp]
\hspace{0.057\linewidth}
    \begin{tabular}{cc}
      \begin{minipage}[t]{0.4\hsize}
        \centering
        %\hspace{0.1\linewidth}
        \includegraphics[width=0.85\linewidth]{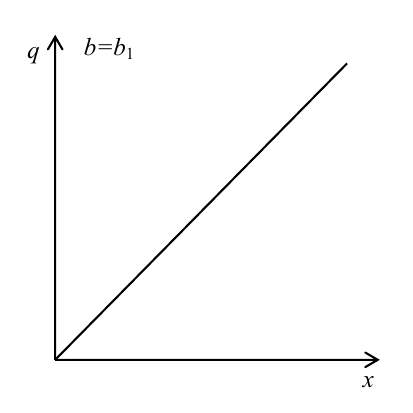}
      \end{minipage} &
      \begin{minipage}[t]{0.4\hsize}
        \centering
        %\hspace{0.1\linewidth}
        \includegraphics[width=0.85\linewidth]{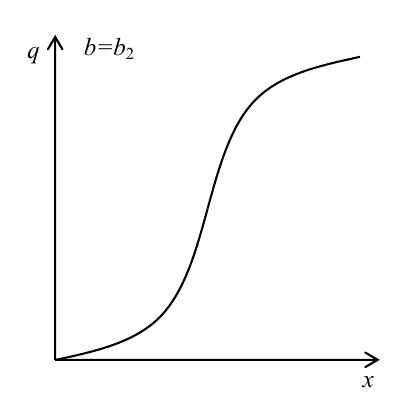}
      \end{minipage} \\
   
      \begin{minipage}[t]{0.4\hsize}
        \centering
        %\hspace{0.1\linewidth}
        \includegraphics[width=0.85\linewidth]{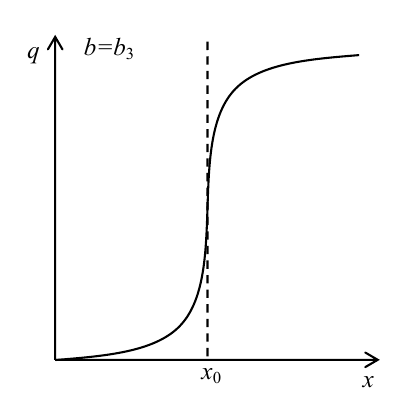}
      \end{minipage} &
      \begin{minipage}[t]{0.4\hsize}
        \centering
        %\hspace{0.1\linewidth}
        \includegraphics[width=0.85\linewidth]{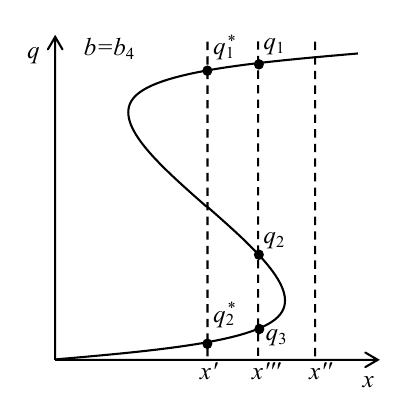}
      \end{minipage} 
    \end{tabular}
    \caption{Graphical explanation of the adhesion model in the case of one dimensional collapse. 
The relation between the Euler coordinate \(x\) and the Lagrange coordinate \(q\) in the Zel'dovich approximation is shown by solid curves.
At a sufficiently small \(b = b_1\) (top left), \(x \approx q\).
At \(b = b_2\) (top right), a little later than \(b_1\), 
the mass moves toward the bottom of the potential under the Zel'dovich approximation, but 
\(q\) is still a single-valued function of \(x\).
At \(b = b_3\) (bottom left), when \({(\partial x(q,b)/\partial q)}_{b}\) become zero
for the first time at \(x=x_0\).
A sheet with infinitesimal mass is formed at this time.
After this time the Zel'dovich approximation is not valid in the collapsed region.
At \(b = b_4>b_3\) (bottom right), the shell-crossing occurs under the Zel'dovich approximation. 
At \(x = x^{\prime \prime}\), \(q\) is a single-valued function of \(x\) and the Zel'dovich approximation is valid.
At \(x = x^{\prime \prime \prime}\), \(q\) is a triple-valued function of \(x\) 
and \(q_1\), \(q_2\) and \(q_3\) have \(x = x^{\prime \prime \prime}\) in the Zel'dovich 
approximation.
In this case, we adopt only \(q_1\) at which \(G(x,q,b)\) takes the maximum value as the relevant solution for Eq.~\eqref{eq:zeldovich}, 
while \(q_2\) and \(q_3\) are absorbed by the sheet collapsed before \(b_4\).
At \(x = x^{\prime}\), there are \(q^*_1\) and \(q^*_2\) at which \(G(x,q,b)\) take the equal maximum value, and all the mass between \(q^*_1\) and \(q^*_2\) is collapsed to a sheet.
The adhesion model mimics the formation of bound systems due to self-gravity, and avoids the unpredictability caused by the shell-crossing in this way. 
}
\label{fig:adhesion}
\end{figure}

For a space-time point specified by \(b\) and \(\bm{x}\) at which we have more than one \(\bm{p}^*\), 
the adhesion model is not identical to the Zel'dovich approximation.
Here, we consider the case in which the maximum is realised for two different values of \(\bm{p}^*\), which we denote by \(\bm{p}^*_1\) and \(\bm{p}^*_2\).
At this space-time point, fluid elements stick together due to the viscosity and a wall with infinitesimal width is formed.
The surface density of the wall is infinite and the normal vector of the wall is proportional to \(\bm{p}_2^* - \bm{p}_1^*\).
\(\bm{x}\) space is divided by the domain walls of the collapsed regions 
and the interior of each domain is described by the Zel'dovich approximation. 
The situation explained above for one-dimensional collapse is explained in Figure~\ref{fig:adhesion}.

If \(\Phi \) is a smooth and bounded potential, at small \(b\) the contribution of the second term of the r.h.s.~in Eq.~\eqref{eq:Gofadhesion} is dominant, 
and hence the number of \(\bm{p}^*\) is one for all \(\bm{x}\).
The Zel'dovich approximation is identical to the adhesion model at small \(b\) for all \(\bm{p}^*\).
Therefore, we can regard \(\bm{p}^*\) as the Lagrange coordinates \(\bm{q}\) for all \( \bm{p}^*\) by using Eq.~\eqref{eq:adhesionxone}, the relation between the Euler coordinates and the Lagrange coordinates at small \(b\).

Here, we consider the properties of the collapsed sheets in the plane-symmetric case.
We set the position of the sheet as \(x = x_{\mathrm{s}}\).
From the mass conservation, the surface density of the sheet, \(\sigma_{\mathrm{s}}\), is given by 
\begin{align}
   \sigma_{\mathrm{s}} =\int ^{x_{\mathrm{s}}+0} _{x_{\mathrm{s}}-0} \rho(x,t) dx = \bar{\rho}(t) (q^*_2 - q^*_1), 
\end{align}
where \(q_1^*\) and \(q_2^*\) are the boundary values of \(q\) of the collapsed region and \(q_2^* - q_1^*\).
Because \(\phi = \frac{3}{2}\dot{a}^2 b G\) in the Zel'dovich approximation from Eq.~\eqref{eq:ZAonephi},
we find that \(\phi\) is continuous across \(x = x_{\mathrm{s}}\), i.e., 
\begin{align}
    \phi(x = x_{\mathrm{s}} + 0) -  \phi(x = x_{\mathrm{s}} - 0) = 0\,.
\end{align}
From Eq.~\eqref{eq:zeldovich} and \eqref{eq:velocity}, the velocity \(v\) (or \(U\)) has a finite gap,
\begin{align}
    v(x = x_{\mathrm{s}} + 0) - v(x = x_{\mathrm{s}} - 0) = -\frac{\dot{b}}{b}(q_1^* - q_2^*).
\end{align}

\subsection{Light propagation}\label{sec:lightpropagation}

Next, we explain the way of estimating the Hubble constant obtained from the directly measurable quantities in the case of the local measurement using the luminosity distance-redshift relation.
In the homogeneous and isotropic universe, 
the Hubble constant is calculated from the luminosity distance \(D_{\mathrm{L}}(z)\) of a light source as a function of the redshift \(z\) of the source.
At a small redshift \(z\), the luminosity distance in the flat FLRW universe is expanded as 
\begin{align}\label{eq:DLexpansionz}
    D_{\mathrm{L}}(z) = {\left(H_0\right)}^{-1} \left[z + \frac{1}{2} \left( 1 - q_0 \right) z^2 - \frac{1}{6} \left( 1 - q_0 - 3 q_0^3 + j_0 \right) z^3 + \cdots \right]\, ,
\end{align}
where \( q_0 \equiv - (H_0)^{-2} d^2 a/dt^2 |_{t=t_0}\) and \(j_0 \equiv (H_0)^{-3} d^3 a/dt^3 |_{t=t_0}\).

However, 
in the actual inhomogeneous universe, 
\(D_{\mathrm{L}}(z)\) is modified.
We compare \(D_{\mathrm{L}}(z)\) in the inhomogeneous universe with the one in 
the homogeneous and isotropic background universe,
in order to evaluate the effects of small-scale inhomogeneities on the measurement of the Hubble constant.
We represent the Hubble constant measured in the background universe and that in the inhomogeneous universe as \(H_{0}^{\mathrm{(B)}}\) and \(H_{0}^{\mathrm{(I)}}\), respectively.
In the universe dominated by a dust fluid \(H_{0}^{\mathrm{(I)}}\) would be defined 
non-linearly as 
\begin{align}\label{eq:H0DLdust}
    H_{0}^{\mathrm{(I)}} \equiv \frac{2}{D_L(z)} \left( 1+z - \sqrt{1+z} \right).
\end{align}
The inhomogeneities of the universe effectively modify not only \(H_0\) but also 
the other parameters such as \(q_0\).
Here, we neglect 
the effective variations of \(q_0\) and \(j_0\) from the background values in deriving the expression~\eqref{eq:H0DLdust}.  
Hence, the estimated values of $H_{0}^{\mathrm{(I)}}$ should be necessarily $z$-dependent. 
This spurious $z$-dependence, strict speaking, should be removed by varying the values of \(q_0\) and \(j_0\). 
Nevertheless, as long as the $z$-dependence of $H_{0}^{\mathrm{(I)}}$ is weak, 
Eq.~\eqref{eq:H0DLdust} should be enough to estimate the order of magnitude of the 
modification of the values of $H_{0}^{\mathrm{(I)}}$. 

We can calculate \(z\) and \(D_L\) by solving the null geodesic from the light source to the observer in the same way as Ref.~\cite{Bolejko:2017fos}.
The wave number vector of the null geodesic,
\begin{align}
    k^{\mu} = \frac{dx^{\mu}}{d\lambda}\,, 
\end{align}
satisfies \(k^{\mu} k_{\mu} = 0\) and \(\ k^{\mu} k_{\nu;\mu} = 0\), where \(\lambda \) is the affine parameter. 
Assuming that both the light source and the observer are moving together with the 
dust fluid, 
the redshift is defined by 
\begin{align} \label{eq:redshift}
    1+z_{\mathrm{s}} = \frac{-u^{\mu} k_{\mu}|_{\mathrm{s}}}{-u^{\mu} k_{\mu}|_{\mathrm{ob}}}\,,
\end{align}
where the quantities labeled by ``ob'' and ``s'' denote the values evaluated 
at the observer and the light source, respectively. 
Here, we renormalize the wave number vector to satisfy  
\begin{align}\label{eq:affinenormalization}
u^{\mu} k_{\mu}|_{\mathrm{ob}} = -1. 
\end{align}

We decompose \(k^{\mu}\) as 
\begin{align}
    k^{\mu} = (- u^{\nu} k_{\nu}) (u^{\mu} + e^{\mu}),
\end{align}
where \(e^{\mu}\) is a vector perpendicular to $u^\mu$, i.e., \(u^{\mu} e_{\mu} = 0\). 
The condition $k^\mu k_\mu=0$ implies $e^{\mu} e_{\mu} = 1$. 
Here, we introduce the induced metric given by  
\begin{align}
    \tilde{h}_{\mu \nu} \equiv g_{\mu \nu} + u_{\mu} u_{\nu} - e_{\mu} e_{\nu}.
\end{align}
Using this metric, we can define the expansion scalar
\begin{align}\label{eq:nullexpansion}
    \tilde{\Theta} \equiv \frac{1}{2} k^{\mu}_{\ ;\mu}\,,
\end{align}
and the shear tensor 
\begin{align}
    \tilde{\sigma}_{\mu \nu} \equiv \tilde{h}_{\mu}^{\alpha} \tilde{h}_{\nu}^{\beta} k_{\beta;\alpha} - \tilde{h}_{\mu \nu} \tilde{\Theta}\,,
\end{align}
of the null geodesic congruence. 
The angular diameter distance $D_{\mathrm{A}}$ satisfies  
\begin{align}\label{eq:da-th}
    \frac{d}{d\lambda} \ln D_{\mathrm{A}} = \tilde{\Theta}\,.
\end{align}
The luminosity distance is related to the angular diameter distance by the relation (see e.g.~\cite{schneider1992gravitational})
\begin{align}\label{eq:da-dl}
    D_{\mathrm{L}} = {(1+z)}^2 D_{\mathrm{A}} .
\end{align}

When we calculate the Hubble constant using the relations~\eqref{eq:redshift}, \eqref{eq:nullexpansion}, \eqref{eq:da-th} and \eqref{eq:da-dl}, 
we need the derivatives of \(k^{\mu}\), which is a little complicated.
Here, we adopt a method to  compute the Hubble constant 
without calculating the derivatives of \(k^{\mu}\), as was done in Ref.~\cite{Bolejko:2017fos}.
The expansion rate of the null geodesic congruence, \(\tilde{\Theta}\), 
satisfies the Raychaudhuri equation
\begin{align} \label{eq:rayc-null}
    \frac{d \tilde{\Theta}}{d\lambda} + \tilde{\Theta}^2 + \tilde{\sigma}^2 = - \frac{1}{2} R_{\mu \nu}k^{\mu} k^{\nu}\,,
\end{align}
where \(\tilde{\sigma}^2 = \tilde{\sigma}_{\mu \nu} \tilde{\sigma}^{\mu \nu}\).
Substituting Eq.~\eqref{eq:da-th} into Eq.~\eqref{eq:rayc-null}, we obtain 
\begin{align} \label{eq:da-eq}
    \frac{\mathrm{d}^2 D_{\mathrm{A}}}{\mathrm{d} \lambda^2} = - \left( \tilde{\sigma}^2 + \frac{1}{2}R_{\mu \nu}k^{\mu} k^{\nu} \right) D_{\mathrm{A}}.
\end{align}
 In the dust dominant universe, using the Einstein's equations, 
 the term including \(R _ {\mu \nu}\) in Eq.~\eqref{eq:da-eq} is expressed as
\begin{align}\label{Einsteineqdustkmu}
    R_{\mu \nu}k^{\mu} k^{\nu} = 8 \pi G \rho {(1+z)}^2 ,
\end{align}
where \(\rho \) is the energy density in the rest flame of the fluid.
Assuming the plane symmetry and focusing on a light ray perpendicular to the planes, 
the null shear \(\tilde{\sigma}\) is zero because of the rotational symmetry in each plane.
Thus, we can calculate the angular diameter distance once we know the energy density and the redshift.

In order to obtain an equation to determine the redshift function, we differentiate Eq.~\eqref{eq:redshift} with respect to \(\lambda \), to get 
\begin{align}\label{eq:z-eq}
    \frac{dz}{d\lambda} =&  -k^{\mu} k^{\nu} u_{\nu;\mu}\,.
\end{align}
We can express \(k^{\mu}\) in terms of \(u^{\mu}\) and \(z\) using Eq.~\eqref{eq:redshift} and the null vector condition of \(k^{\mu}\).
Thus, we can solve for the redshift, once we know the expressions for 
\(u_{\nu;\mu}\) and \(u^{\mu}\).
We solve Eqs.~\eqref{eq:da-eq} and~\eqref{eq:z-eq} from the observer to the source 
with the initial conditions, 
\begin{align}
    D_{\mathrm{A}} |_{z =0} &= 0 \, ,\\
\frac{dD_\mathrm{A}}{d\lambda}\biggr|_{z=0} &= 1 \,. \label{eq:initialdDAdlamda}
\end{align}

\subsection{Setup of the model and the error control}
Following the method explained in Section~\ref{sec:lightpropagation}, 
we calculate \(D_{\mathrm{L}}(z)\) and $z$. 
We first approximately evaluate the quantities that are needed to solve Eqs.~\eqref{eq:da-eq} and~\eqref{eq:z-eq} in Newtonian cosmology, 
explaining the order of magnitude that we neglect.

As a simple example of the plane-symmetric case, 
we consider the velocity potential in Eq.~\eqref{eq:zeldovich} given by 
\begin{align}\label{eq:Phiprofile}
    \Phi = -AL^2\exp{\left(- \frac{q^2}{2L^2}\right)},
\end{align}
where \(A\) is the amplitude of the perturbation 
and \(L\) is the comoving length scale of the perturbation.
The amplitudes of perturbation at the observer and the light source are 
so small that they are negligible at those points.
We choose \(b = a/a_{\mathrm{eq}}\) by appropriately rescaling \(\Phi\), because \(b\) and \(\Phi \) always 
appear as a product in the physical quantities.
Then, \(A\) unambiguously determines the amplitude of the perturbation.

To evaluate the order of magnitude neglected in our calculations, we denote \(\rho \), \(v\) and \(\phi \) together by \(\alpha \) and we 
decompose \(\alpha \) as \(\alpha = \alpha^{\mathrm{(B)}} + \delta \alpha\), where $\alpha^{\mathrm{(B)}}$ is the background quantity. 
In addition, we expand \(\delta \alpha\) as \( \delta \alpha = \alpha^{(1)}+ \alpha^{(2)} + \cdots \), 
where $\alpha^{(1)}$ describes 
the leading order perturbation. Hence, $\rho^{(1)}/\rho^{(\mathrm{B})}$, $v^{(1)}$ and $\phi^{(1)}$ are $\mathcal{O}(\epsilon^0)$, $\mathcal{O}(\epsilon)$ and 
$\mathcal{O}(\epsilon^2)$, respectively. 
We find that the higher order terms scale like 
\(\alpha^{(2)} \sim \alpha^{(1)} \mathcal{O}(\epsilon^2)\).
The equations of Newtonian cosmology, Eqs.\eqref{eq:NGeq}-\eqref{eq:NGfluidEulereq} can have corrections higher order in \(\epsilon \).
Therefore, they are not sufficient to determine \(\alpha^{(2)}\). 
The order of magnitude of the errors in the computation based on Newtonian cosmology is evaluated by estimating the terms that contain \(\alpha^{(2)}\). 
From this discussion, we may neglect the terms even if they 
are composed of \(\alpha^{(1)}\), 
in case they are as small as the terms that depend on \(\alpha^{(2)}\).

We solve Eqs.~\eqref{eq:da-eq} and~\eqref{eq:z-eq} from \(z = 0\) to \(z = z_{\mathrm{s}}\), where \(z_{\mathrm{s}}\) is the redshift of the source.
We choose the source location so that \(z_{\mathrm{s}} = 0.1\).
Thus, for the physical quantities along the light ray, we can 
use the approximate relation 
\begin{align} \label{eq:scalefactorcounting}
    a(t) = \mathcal{O}(1),
\end{align}
for the order of magnitude counting.
From \(k^{\mu} k_{\mu} = 0\) and Eq.~\eqref{eq:redshift}, we find 
\begin{align} \label{eq:wavevectorcounting}
    k^{t\mathrm{(B)}} \sim k^{x\mathrm{(B)}} = \mathcal{O}(1)\,,
\end{align}
where \(k^{\mu\mathrm{(B)}}\) is the wave number vector \(k^{\mu}\) for the background universe.
We show that the relative errors neglected in calculating the Hubble constant are $\mathcal{O}(H\lambda_{\mathrm{s}} \epsilon^3)$, where \(\lambda_{\mathrm{s}}\) is the distance between the observer and the source. 
In evaluating the errors in the Hubble constant in the inhomogeneous universe, 
we examine the order of magnitude of the leading order contribution coming from \(\alpha^{(2)}\). 
\(D_{\mathrm{L}}(z_{\mathrm{s}})\) can be decomposed as \(D_{\mathrm{L}}(z_{\mathrm{s}}) = D_{\mathrm{L}}^{\mathrm{(B)}}(z_{\mathrm{s}}) + \delta D_{\mathrm{L}}(z_{\mathrm{s}})\), where the superscript \(\mathrm{(B)}\) represents the background quantity and \(\delta D_{\mathrm{L}}\) is the remaing part of the luminosity distance \(D_{\mathrm{L}}\) coming from the perturbations.
Then, we find 
\begin{align}
    \frac{H_{0}^{\mathrm{(I)}} - H_0^{\mathrm{(B)}}}{H_0^{\mathrm{(B)}}} = \frac{\frac{2}{D_{\mathrm{L}}(z_{\mathrm{s}})} \left( 1+z_{\mathrm{s}} - \sqrt{1+z_{\mathrm{s}}} \right)}{\frac{2}{D_{\mathrm{L}}^{\mathrm{(B)}}(z_{\mathrm{s}})} \left( 1+z_{\mathrm{s}} - \sqrt{1+z_{\mathrm{s}}} \right)} -1
    \simeq - \frac{\delta D_{\mathrm{L}}(z_{\mathrm{s}})}{D_{\mathrm{L}}^{\mathrm{(B)}}(z_{\mathrm{s}})}, 
\end{align}
where we have assumed \(\delta D_{\mathrm{L}}/D_{\mathrm{L}}^{\mathrm{(B)}} \ll 1\).
Therefore, we need to estimate the error in the estimate of \( \delta D_{\mathrm{L}}(z_{\mathrm{s}})/D_{\mathrm{L}}^{(\mathrm{B})}(z_{\mathrm{s}})\).

To evaluate $\delta D_{\mathrm{L}}(z_{\mathrm{s}})$, we first consider \(k^\mu \).
From Eqs~\eqref{eq:4velocity0},~\eqref{eq:4velocityi} and~\eqref{eq:redshift}, \(k^{\mu}\) is expressed as 
\begin{align}
    k^t & = \left(1 - \phi + \frac{1}{2} a^2 v^2 - av \right) (1 + z) + \mathcal{O}(\epsilon^3), \label{eq:kt}\\
    a k^x & = - \left( 1 + \phi + \frac{1}{2} a^2 v^2 - av \right) (1+z) + \mathcal{O}(\epsilon^3), \label{eq:kx}
\end{align}
where $\mathcal{O}(\epsilon^3)$ in \(k^t\) and \(k^x\) corresponds to the leading order error coming from  \(v^{(2)}\) and \(z\).\footnote{When we consider to directly solve the null geodesic equations, the relative error of \(k^{\mu}\) coming from \(\alpha^{(2)}\) is estimated 
to be \(\mathcal{O}(\epsilon^4)\), which is even higher order.}
We assume the error of \(z\) coming from \(\alpha^{(2)} \) is of \(\mathcal{O}(\epsilon^3)\).

Next, we consider the redshift \(z\).
In order to solve Eq.~\eqref{eq:z-eq}, we need the expression for \(u_{b;a}\).
From Eqs.\eqref{eq:4velocity0} and~\eqref{eq:4velocityi}, the components of \(u_{b;a}\) are calculated as 
\begin{align}
    u_{t;t} &= L^{-1}\mathcal{O}(\varepsilon^3)\,,
    \label{eq:utt}  \\
    u_{t;x} &= -a^2v \left({\left(\frac{\partial v}{\partial x}\right)}_{t} + H\right) 
   +L^{-1}\mathcal{O}(\varepsilon^4)\,,\\
    u_{x:t} &=  Ha^2v + a^2 {\left(\frac{\partial v}{\partial t}\right)}_{x} + {\left(\frac{\partial \phi}{\partial x}\right)}_{t}
   +L^{-1}\mathcal{O}(\varepsilon^4)\,,\\
    u_{x;x} &= a^2 {\left(\frac{\partial v}{\partial x}\right)}_{t} + a^2 H (1 - 3 \phi)
   +L^{-1}\mathcal{O}(\varepsilon^3)\,, \label{eq:uxx}
\end{align}
where the leading order of the neglected terms in \(u_{\mu ; \nu}\) are $\partial v^{(2)}/\partial x =H\mathcal{O}(\epsilon^2)=L^{-1}\mathcal{O}(\epsilon^3)$ in $u_{t;t}$ and $u_{x;x}$. 
Although the term including $\phi$ in the factor \((1 - 3 \phi )\) in Eq.~\eqref{eq:uxx} is as small as 
the order of magnitude that we neglect, 
we keep it here, since the term remains even in the linear perturbation theory. 
We should recall that $\phi\gg v^2$ for the large scale perturbation, although we count both of them as $\mathcal{O}(\epsilon^2)$. 
We neglect the terms including \({\left(\partial \phi/\partial t\right)}_{x}\) 
because \(\left(\partial v^{(2)}/\partial x \right)_{t}\) is the same order and \({\left(\partial \phi/\partial t\right)}_{x}\) vanishes in the linear approximation.

In estimating the error of \(z\) from Eq.~\eqref{eq:z-eq}, we also need to consider the perturbation of the affine parameter, which can be expressed as \(\lambda(t,q) = \lambda^{\mathrm{(B)}}(t,q) + \delta \lambda(t,q)\).
From Eq.~\eqref{eq:kt} and \eqref{eq:kx}, the terms in \(k^{\mu}\) dependent on \(\alpha^{(2)}\) lead to the error in \(\delta \lambda\) of  \(\mathcal{O}(H^{-1} \epsilon^4)\).

The redshift can be decomposed as \(z(t,q) = z^{\mathrm{(B)}}(t,q) + \delta z(t,q)\).
From Eq.~\eqref{eq:z-eq}, the terms in \(u_{\nu;\mu}\) dependent on \(\alpha^{(2)}\) lead to the error in \(\delta z\) of \(\mathcal{O}(\epsilon^3)\).
The terms coming from \(\alpha^{(2)}\) in \(k^{\mu}\) and \(\lambda\) also produce errors in \(\delta z\), 
but they are even higher order. 
Therefore, we conclude that the magnitude of the neglected higher order terms in \(\delta z\) is of \(\mathcal{O}(\epsilon^3)\) and this is consistent with the above assumption of the error of \(z\).

Now, we are ready to consider \(\delta D_{\mathrm{A}}\). 
The leading order term coming from \(\alpha^{(2)}\) in \(\delta D_{\mathrm{A}}(z_{\mathrm{s}})\) is due to the terms of \(\mathcal{O}(\epsilon^2)\) in \(\rho^{(2)}/\rho^{(\mathrm{B})}\). 
From Eqs.~\eqref{eq:da-eq}, these terms produce 
\begin{align}
    \mbox{error in }\delta D_{\mathrm{A}} = \mathcal{O}(H\lambda^2_{\mathrm{s}}\epsilon^3)\,.
\end{align}
Therefore, this effect produces an error in the ratio \( {\delta D_{\mathrm{A}}(z_{\mathrm{s}})}/{D_{\mathrm{A}}^{(B)}(z_{\mathrm{s}})}\) of  \(\mathcal{O}(H\lambda_{\mathrm{s}} \epsilon^3 )\), where $H\lambda_{\mathrm{s}} = \mathcal{O}(10^{-1})$ in our case.
From Eq.~\eqref{eq:da-eq}, we find that the leading order terms coming from \(\alpha^{(2)}\) in \(\delta z\) and \(\delta \lambda\) produce 
only the higher order errors in \({\delta D_{\mathrm{A}}(z_{\mathrm{s}})}/{D_{\mathrm{A}}^{(B)}(z_{\mathrm{s}})}\).  
From the above, we find that the order of magnitude of the error in our estimate of the Hubble constant is the error in \((H_{0}^{\mathrm{(I)}} - H_0^{\mathrm{(B)}})/H_0^{(B)}\), which is of \(\mathcal{O}(H\lambda_{\mathrm{s}}\epsilon^3) \).
Therefore, the Newtonian model is precise enough to discuss the Hubble tension neglecting relative error of \(\mathcal{O}(H\lambda_{\mathrm{s}}\epsilon^3)\).
Here, we consider a single over-density region,
but even if we consider a sequence of over-density regions, we only need to multiply the factor \(\lambda_{\mathrm{s}}/L \) to the above estimate of the error, 
and hence it is still precise enough to discuss the Hubble tension since typically \(\epsilon^2 \simeq 10^{-6}\) or so.
This would mean that any effort to explain the Hubble tension by means of the 
non-linear effect beyond Newtonian cosmology should fail at the level of order counting. 

\subsection{Hubble constant evaluation}
Here, we explain how we solve Eqs.~\eqref{eq:da-eq} and~\eqref{eq:z-eq} in the adhesion model.
First, we consider the space-time region \((x, t)\) where \(q^*\), the point in Lagrange coordinates that maximizes \(G(x,q,b)\), is unique.
In order to solve Eq.~\eqref{eq:da-eq}, we need \(\rho \), the density of the fluid in the rest flame.
In this region, the Zel'dovich approximation is valid.
Therefore, \(\rho \) is given by Eq.~\eqref{eq:density-newton}.
In order to solve Eq.~\eqref{eq:z-eq}, we use Eqs.~\eqref{eq:kt}-\eqref{eq:uxx}.
\(v\) and \(\phi \) are obtained by 
Eqs.~\eqref{eq:velocity} and~\eqref{eq:ZAonephi}, respectively.

After caustics formation, sheet-like collapsed regions are formed.
In solving Eqs.~\eqref{eq:da-eq} and~\eqref{eq:z-eq} across the collapsed region, 
we need to take into account the singular distribution of the matter energy density.  
In the present model we have, at most, only one sheet of collapsed region. 

First, we consider the junction condition of \(k^{\mu}\).
In the $(t,x)$ coordinates, the metric and its derivatives are finite. 
Therefore, we find that \(k^{\mu}\) must be continuous at the position of the sheet, \(x = x_{\mathrm{s}}\), from the geodesic equation.

Next, we express the junction conditions for Eqs.~\eqref{eq:da-eq} and~\eqref{eq:z-eq}.
From the form of \(\Phi(q)\) given in Eq.~\eqref{eq:Phiprofile}, 
the number of \(q^*\) 
is two, 
only at \(x = 0\) and we set \( x_{\mathrm{s}} = 0 \) from here on. 
At \(x = 0\), $u^0$ is continuous due to the reflection symmetric, while $u^x\sim v$ is not. 
Hence, from Eq.~\eqref{eq:redshift}, we have 
\begin{align}    {\left[ z \right]}^{x=+0} _{x=-0} 
    = k_{x}{[v]}^{x=+0} _{x=-0}, 
\end{align}
where \( {\left[ \cdots \right]}^{x=+0} _{x=-0} \) represents the difference between the values in the square brackets evaluated before and after the jump at \(x = 0\).
For the angular diameter distance \(D_{\mathrm{A}}\), we can integrate Eq.\eqref{eq:da-eq} 
in the neighborhood of \(x = 0\) to find 
\begin{align}
    {\left[\frac{d D_{\mathrm{A}}}{d \lambda}\right]}^{x = +0}_{ x = -0} &= - 4 \pi G D_{\mathrm{A}} \int^{\lambda(x = +0)}_{\lambda(x = -0)} \rho {(1 + z)}^2 d \lambda \nonumber \\
    & = - 4 \pi G D_{\mathrm{A}} \frac{{(k^t)}^2}{k^x} \int^{x = +0}_{x = -0} \rho dx \nonumber \\
    & = - 4 \pi G D_{\mathrm{A}}\frac{{(k^t)}^2}{k^x} \bar{\rho}(t) (q_2^* - q_1^*), \label{eq:DAdlambdagap}
\end{align}
where we have used the fact that the velocity of the sheet at \(x = 0\) is zero.  

In the following explicit calculation, we set \(L = 10 \,\mathrm{Mpc}\) as a model of a cluster of galaxies.
We put the perturbation only in the middle between the observer and the light source 
so that the amplitudes of perturbation evaluated on both ends of light ray are sufficiently small.
We set \(H_0^{\mathrm{(B)}} = 67 \,  \mathrm{km/s/Mpc}\) and locate the observer at \((t,q) = (t_0 = 2/(3H_0^{\mathrm{(B)}}),100 \,\mathrm{Mpc})\).
\begin{figure}[tbp]
    \centering
    \includegraphics[width=0.49\textwidth]{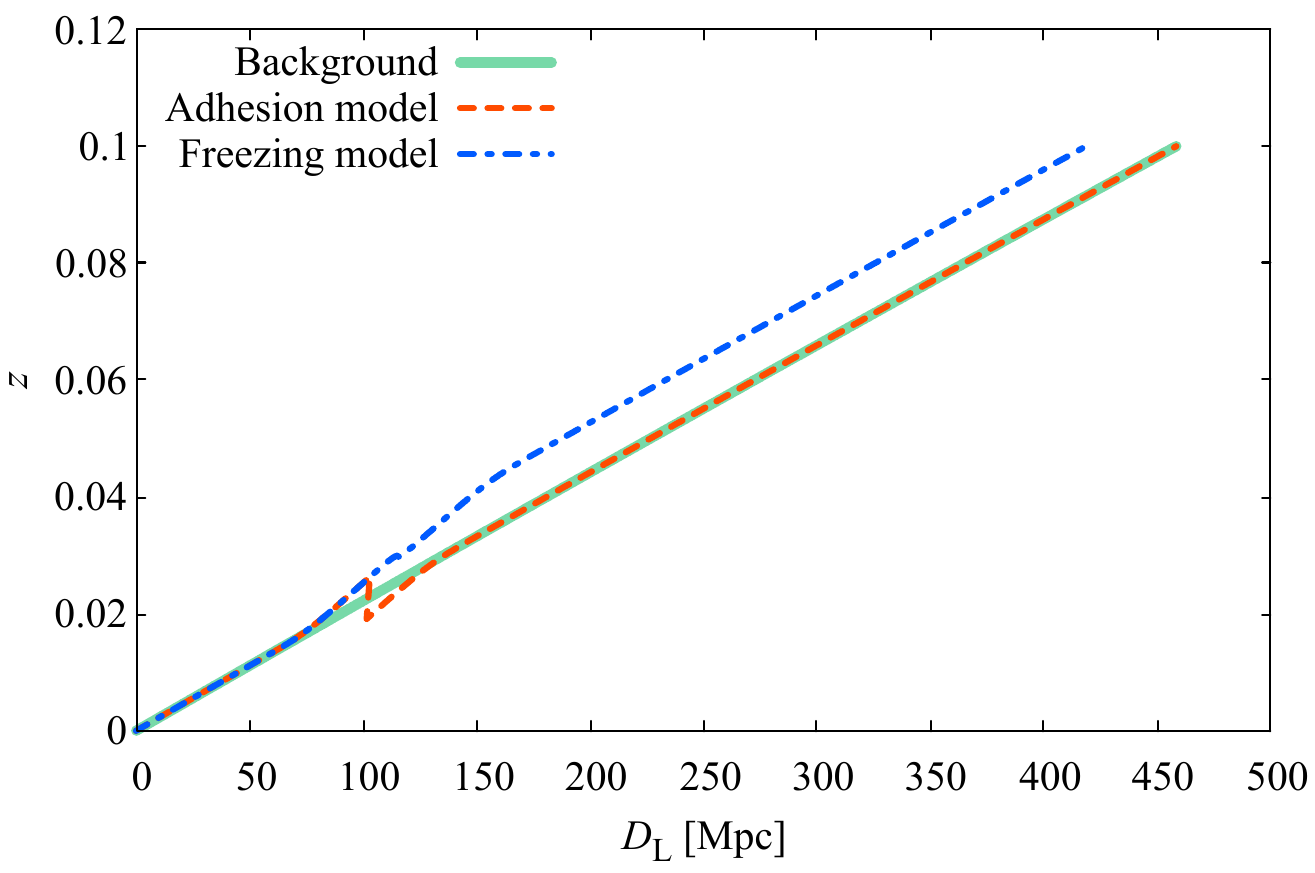} 
    \caption{The luminosity distance-redshift relations in the model used in this work in the case of \(A = 10^{-3}\).
    The red dashed curve denotes the one evaluated using the adhesion model described in Section~\ref{sec:h0estimation} and the blue dash-dotted curve denotes the one for the freezing model explained in Section~\ref{sec:freezing}.
    The green solid line denotes the case of \(A = 0\), i.e., the reference background universe.}\label{fig:DLz}
\end{figure}
We calculate the luminosity distance in the case of \(A = 10^{-3}\), corresponding to \(|\phi(t=0,q=0)| = 2.6 \times 10^{-5}\).
In Figure~\ref{fig:DLz}, we show the Hubble diagram, i.e., the relation between the redshift and the luminosity distance.
The green solid line shows the case with \(\Psi(q) = 0\), which corresponds to the reference homogeneous isotropic universe. 
The red dashed curve shows the dependence in the case of the adhesion model. 
The trajectory in this Hubble diagram is largely modified from the green solid line around the collapsed region.  
The jump in $z$ is caused by the sudden change of the matter velocity across the sheet, to which the matter is accreting. 
Nevertheless, the Hubble constant determined by $D_L$ at the source with the redshift \(z_{\mathrm{s}}\) remains unchanged. 
This is because the effect of inhomogeneities, i.e., the difference between the green solid line and the red dashed curve is mainly coming from the peculiar motion of the matter. 
In fact, the numerically evaluated relative difference between \(H_0^{\mathrm{(I)}}\) and \(H_0^{\mathrm{(B)}}\) is as small as 
\begin{align}
\label{eq:relativediff}
\delta_{\mathrm{rel}, \, H_0} \equiv \frac{|H_{0}^{\mathrm{(I)}} - H_0^{\mathrm{(B)}}|}{H_0^{\mathrm{(B)}}}\simeq 2.7 \times 10^{-6}\,.
\end{align}
Thus, the effect of inhomogeneities on the measured Hubble constant in our model is small, well below the accuracies of the present measurements and  the magnitude that would explain the Hubble tension.

\section{Freezing model case}\label{sec:freezing}
It is convenient to use analytic models in order to reduce the computational cost or to obtain an intuitive understanding.
In Section~\ref{sec:h0estimation}, we discussed the treatment of gravitationally collapsed regions using the adhesion model. 
In this section we will give a lesson that we must be careful in the treatment of the gravitationally collapsed regions in 
investigating the effects of inhomogeneities using analytic models. 
For this purpose, we demonstrate how an alternative simple treatment can lead to an unphysical conclusion.
The prescription discussed in this section as an example 
is a model in which the evolution of the expansion rate, density and shear of the fluid element are frozen 
after passing the point where the expansion rate becomes zero,
which was used in~\cite{Bolejko:2017wfy}.
We will discuss the details later.
Here, we use the synchronous and comoving coordinates.
The synchronous and comoving coordinates are often used because the perturbation equations become easy to handle, especially for the dust dominant universe.
We derive the metric in the synchronous and comoving coordinates transformed from the metric in the Newtonian gauge under the Zel'dovich approximation.
We should adopt $q$ as the spatial coordinate in the synchronous and comoving gauge. 
From~\eqref{eq:zeldovich}, we get 
\begin{align}\label{eq:zeldovichdif}
    dx = \left( 1 - b(t)\frac{d^2 \Phi}{dq^2}(q)\right) dq - \dot{b}(t) \frac{d \Phi}{dq}(q) dt
    ={\left(\frac{\bar{\rho}(t)}{\rho(t,q)}\right)}dq+v(t,q) dt\,.
\end{align}
Substituting this relation 
into the metric in the Newtonian gauge~\eqref{eq:newtoniangauge}, we obtain 
\begin{align}\label{eq:intervaltq}
    ds^2 = &- \left\lbrace 1 + 2 \phi(t,q) - a^2(t) v^2(t,q) \right\rbrace  dt^2 + 2 a^2(t) v(t,q) \frac{\bar{\rho}(t)}{\rho(t,q)} dt dq \nonumber \\
    &+ a^2(t) (1 - 2 \phi(t,q)){\left(\frac{\bar{\rho}(t)}{\rho(t,q)}\right)}^2 dq^2 + a^2(t) (1 - 2 \phi(t,q)) (dy^2 + dz^2),
\end{align}
where we have neglected the terms of the orders of corrections from $\alpha^{(2)}$. 
The factor \((1 - 2 \phi(t,q) )\) in the components of \(dq^2\) can be reduced to 1 in comparison with the order of magnitude that we neglect, recalling that $\rho^{(2)}/\rho^{(\mathrm{B})}=\mathcal{O}(\varepsilon^2)$. 
Nevertheless, we keep this factor as before, since it appears in the linear perturbation theory. 
To change the time coordinate to achieve the synchronous condition, we introduce a new time coordinate, 
\begin{align}
    T = (1 + \phi(t,q))t, 
\end{align}
and then, we find 
$$
 dT=(1+\phi(t,q)- \frac{1}{2} a^2(t) v^2(t,q))dt -a^2(t) v(t,q) \frac{\bar\rho(t)}{\rho(t,q)}dq\,.
$$
Using this time coordinate, we find 
\begin{align}\label{eq:synchronouscomoving}
    ds^2 =& - dT^2 + a^2(T) \left(1 - \frac{10}{3} \phi(T,q)\right) {\left( 1 - b(T) \frac{d^2 \Phi}{d q^2} \right)}^2 dq^2 
    \cr 
    & + a^2(T) \left(1 - \frac{10}{3} \phi(T,q)\right) (dy^2 + dz^2)\,, 
\end{align}
where we again omit the terms higher order in $\epsilon$ and use the relation $a(t)\approx (1 - 2\phi/3 )a(T)$. 

In the synchronous and comoving coordinates \((T, q, y, z)\), the dust fluid 4-velocity is 
given by 
\begin{align}
    u^T = 1, \quad u^q = u^y = u^z = 0.
\end{align}
In these coordinates, the expansion of the dust fluid \(\Theta \) is expressed as 
\begin{align}
    \Theta &= u^{\mu}_{\,;\mu} = \Gamma^\mu_{T \mu} = \frac{1}{2}\frac{\gamma_{qq,T}}{\gamma_{qq}} + \frac{\gamma_{yy,T}}{\gamma_{yy}} \nonumber \\
      &= 3 H(T) - \frac{b^{\prime}(T)\frac{d^2 \Phi}{dq^2}}{1 - b(T) \frac{d^2 \Phi}{dq^2}},
\end{align}
where \(\gamma_{ij}\) represents the spatial metric for the line element in 
Eq.~\eqref{eq:synchronouscomoving} and \(b^{\prime} \equiv {d b}/{d T}\).
The non-vanishing elements of the shear tenser are expressed as 
\begin{align}
    \sigma_{qq} &= \frac{1}{3} \gamma_{qq} \left(  \frac{\gamma_{qq,T}}{\gamma_{qq}} - \frac{\gamma_{yy,T}}{\gamma_{yy}} \right) = -2 \gamma_{qq}\Sigma\,, \\
    \sigma_{yy} &= \sigma_{zz} = \frac{1}{6} \gamma_{yy} \left(  \frac{\gamma_{yy,T}}{\gamma_{yy}} - \frac{\gamma_{qq,T}}{\gamma_{qq}} \right) = \gamma_{yy}\Sigma\,,
\end{align}
where 
\begin{align}
\Sigma \equiv \frac{1}{6} \left(  \frac{\gamma_{yy,T}}{\gamma_{yy}} - \frac{\gamma_{qq,T}}{\gamma_{qq}} \right)
 = \frac{1}{3} \frac{b^{\prime}(T)\frac{d^2 \Phi}{dq^2}}{1 - b(T) \frac{d^2 \Phi}{dq^2}}\,.
\end{align}
The density given in Eq.~\eqref{eq:density-newton} is rewritten in terms of these coordinates as 
\begin{align}
    \rho=\rho(t(T,q),q)= \frac{\bar{\rho}(T)(1 + 2 \phi(T,q))}{1 - b(T)\frac{d^2 \Phi}{dq^2}}\,,
\end{align}
neglecting the higher order terms. The factor $(1+2\phi(T,q))$ arises from $\bar\rho(t)=(1+2\phi)\bar\rho(T)$ and kept here because the factor is present in the linear perturbation theory.

As we mentioned earlier, the Zel'dovich approximation breaks down when caustics form.
In this section, for the prescription to avoid the caustic formation, 
we freeze the values of \(\Theta \), \(\Sigma \) and \(\rho \)
 after the time \( T_{\mathrm{FRZ}}\) when the expansion \(\Theta\) vanishes. 
Then, the metric for $T>T_{\mathrm{FRZ}}$ is expressed as 
\begin{align}
    \gamma_{qq} (T,q) &= \gamma_{qq} (T_{\mathrm{FRZ}},q) \exp \left\lbrace -4 \Sigma(T_{\mathrm{FRZ}},q) (T - T_{\mathrm{FRZ}})\right\rbrace, \\
    \gamma_{yy} (T,q) &= \gamma_{zz}(T,q) = \gamma_{yy} (T_{\mathrm{FRZ}},q) \exp \left\lbrace 2 \Sigma(T_{\mathrm{FRZ}},q) (T - T_{\mathrm{FRZ}})\right\rbrace.
\end{align}
In this freezing treatment
the fluid elements contract in \(x\) direction and expand in \(y\) and \(z\) directions for \(T > T_{\mathrm{FRZ}}\).  
It is convenient, using \(\Theta \) and \(\Sigma \), to rewrite Eq.~\eqref{eq:z-eq} in the freezing model as 
\begin{align}\label{eq:z-eq2}
    \frac{dz}{d \lambda} = - \left( -2 \Sigma + \frac{1}{3} \Theta \right) {(1 + z)}^2.
\end{align}

\begin{figure}[tbp]
    \centering
        \includegraphics[width=0.49\textwidth]{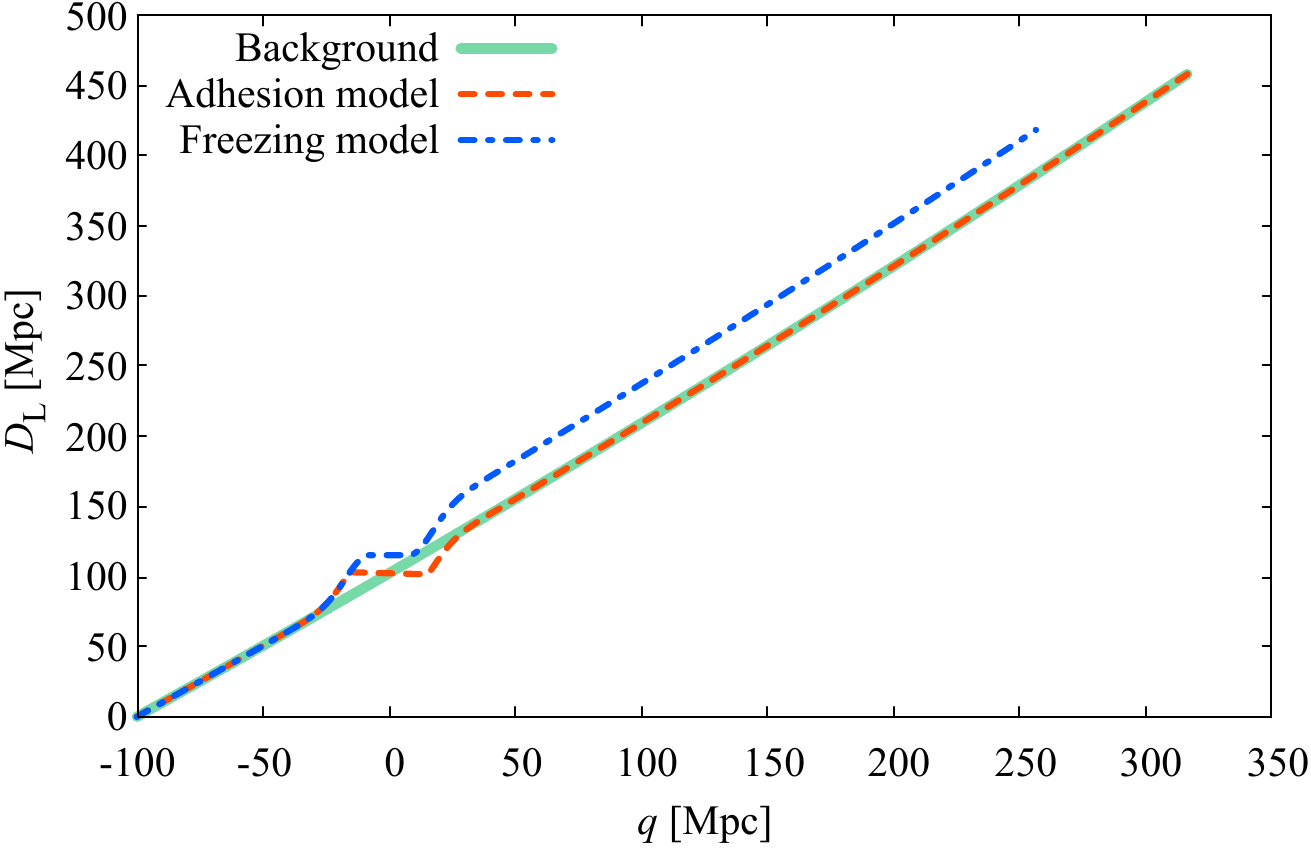} 
        \includegraphics[width=0.49\textwidth]{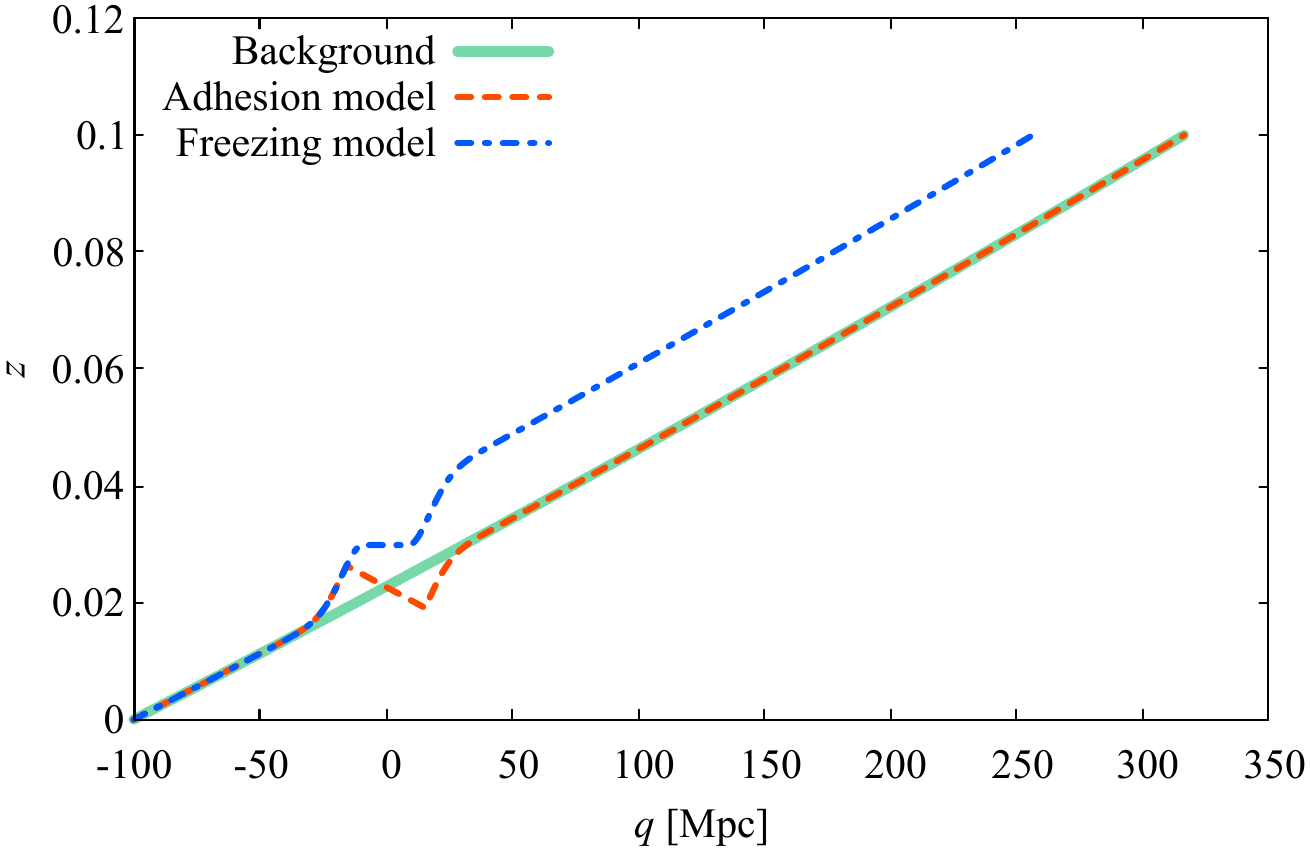} 
        \caption{
        Left: The luminosity distances from the source as a function of  the Lagrange coordinate \(q\) in the models used in this work in the case of \(A = 10^{-3}\).
        The red dashed curve denotes the one for the adhesion model discussed in Section~\ref{sec:h0estimation} and the blue dash-dotted curve denotes the one for the freezing model introduced in Section~\ref{sec:freezing}.
        The green solid line denotes the case of \(A = 0\), i.e., the reference background universe.
        Right: The plot similar to the left panel but for 
        the redshifts of the source.
        }\label{fig:qDLandqz}
    \end{figure}

The blue dash-dotted curve in Figure~\ref{fig:DLz} shows the results obtained by applying the 
freezing treatment. 
We find that the Hubble constant becomes \(H_{0}^{\mathrm{(I)}} \simeq 73.4 \mathrm{km/s/Mpc} \), 
which is much larger than the background value.
The relative difference defined in Eq.~\eqref{eq:relativediff} is as large as \(\delta_{\mathrm{rel}, \, H_0} \simeq 9.5 \times 10^{-2}\).
In the left panel of Figure~\ref{fig:qDLandqz}, we show the dependence of the luminosity distance on the Lagrange coordinate $q$.  
Also, in the right panel of Figure~\ref{fig:qDLandqz}, we show the dependence of the redshift on the Lagrange coordinate.
You can see the two discontinuities of the derivative around \(q=0\) in Figure~\ref{fig:qDLandqz}  in the adhesion model.
The fluids which have the Lagrange coordinates between these two discontinuities are all corresponding to \(x=0\).
As shown in Section~\ref{sec:h0estimation}, \(D_L\) is continuous while \(z\) has a gap inherited from the discontinuous velocity field at \(x=0\) in the adhesion model.

In the freezing model, the flat regions are corresponding to the frozen region in Figure~\ref{fig:qDLandqz}.
The reason for the difference from the background value at a large \(q\) is simply because the delicate cancellation of the effects of inhomogeneity is not 
appropriately treated in this model in contrast to the adhesion model. 
\begin{figure}[tbp]
    \centering
        \includegraphics[width=0.49\textwidth]{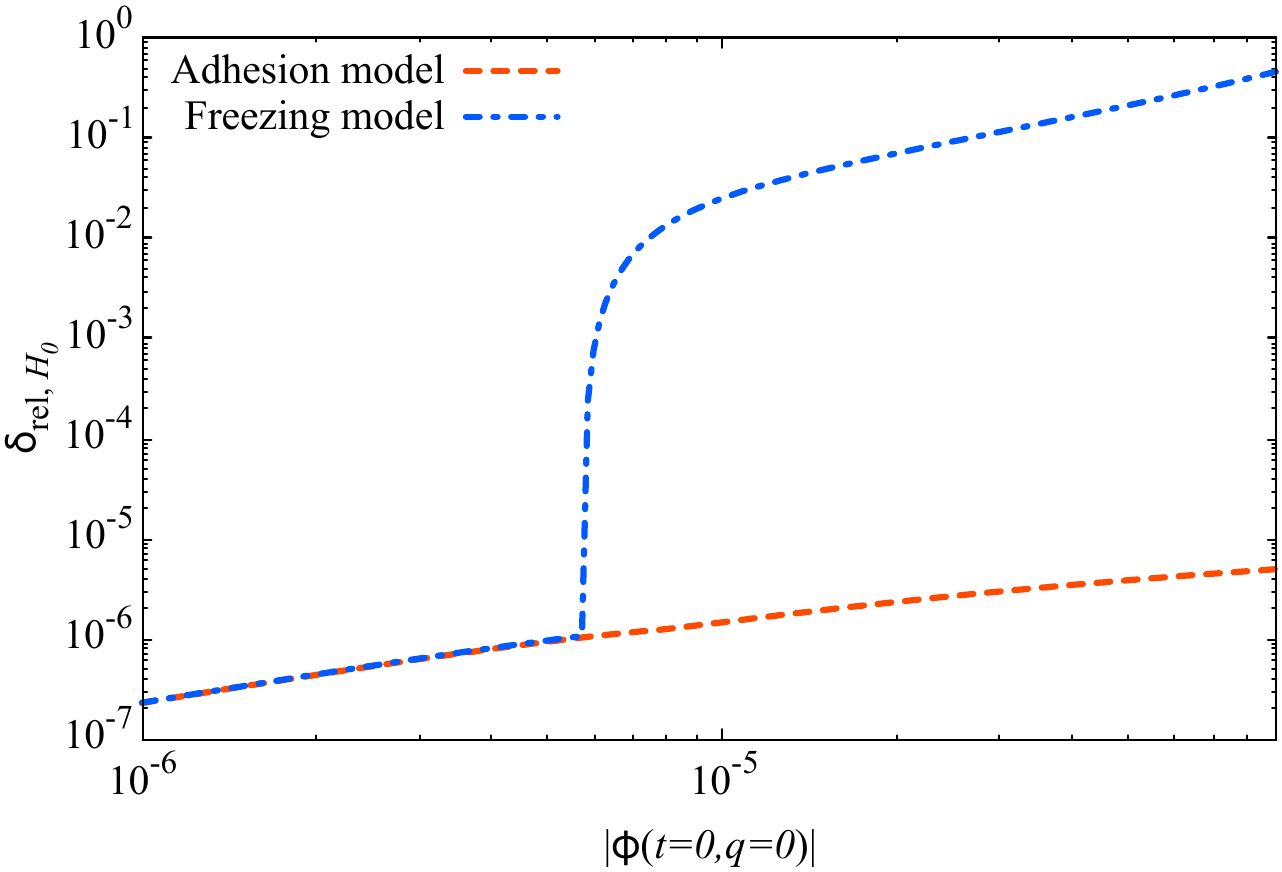}
        \caption{
        The dependence of the error on the amplitude of the velocity potential. 
        \(\delta_{\mathrm{rel}, \, H_0} \equiv \sqrt{{\{(H_{0}^{\mathrm{(I)}} - H_0^{\mathrm{(B)}})/H_0^{\mathrm{(B)}}\}}^2}\) is the deviation of the measured values of the Hubble constant from that of the background, 
        where \(H_0^{\mathrm{(B)}}\) is the Hubble constant of the background.
        The red dashed and blue dash-dotted curves are the results for the adhesion model and the freezing model, respectively.
        }\label{fig:delta_relH0}
\end{figure}

In Figure~\ref{fig:delta_relH0}, we show $\delta_{\mathrm{rel}, \, H_0}$ as a function of the initial amplitude of the potential, \(|\phi(t=0,q=0)|\). 
In the adhesion model, which is drawn by the red dashed curve, the relative difference between \(H_0^{\mathrm{(B)}}\) and \(H_{0}^{\mathrm{(I)}}\) remains small. 
In the adhesion model, for \(|\phi(t=0,q=0)| \gtrsim 8 \times 10^{-6}\), the collapse occurs sufficiently quickly, and the light ray passes through a sheet-like collapsed region. 
Nevertheless, $\delta_{\mathrm{rel}, \, H_0}$ smoothly varies depending 
on the initial amplitude of the potential.
On the other hand, in the freezing model, which is shown by the blue dash-dotted curve, 
 $\delta_{\mathrm{rel}, \, H_0}$ remains to be the same as the adhesion model 
 for \(|\phi(t=0,q=0)| \lesssim 6 \times 10^{-6}\),
but $\delta_{\mathrm{rel}, \, H_0}$ rapidly starts to increase 
as the amplitude of the potential exceeds the critical value. 
Namely, once the frozen region starts to affect the light propagation, the error caused by the inappropriate prescription in the 
freezing model becomes extremely large.
    \begin{figure}[tbp]
    \centering
        \includegraphics[width=0.49\textwidth]{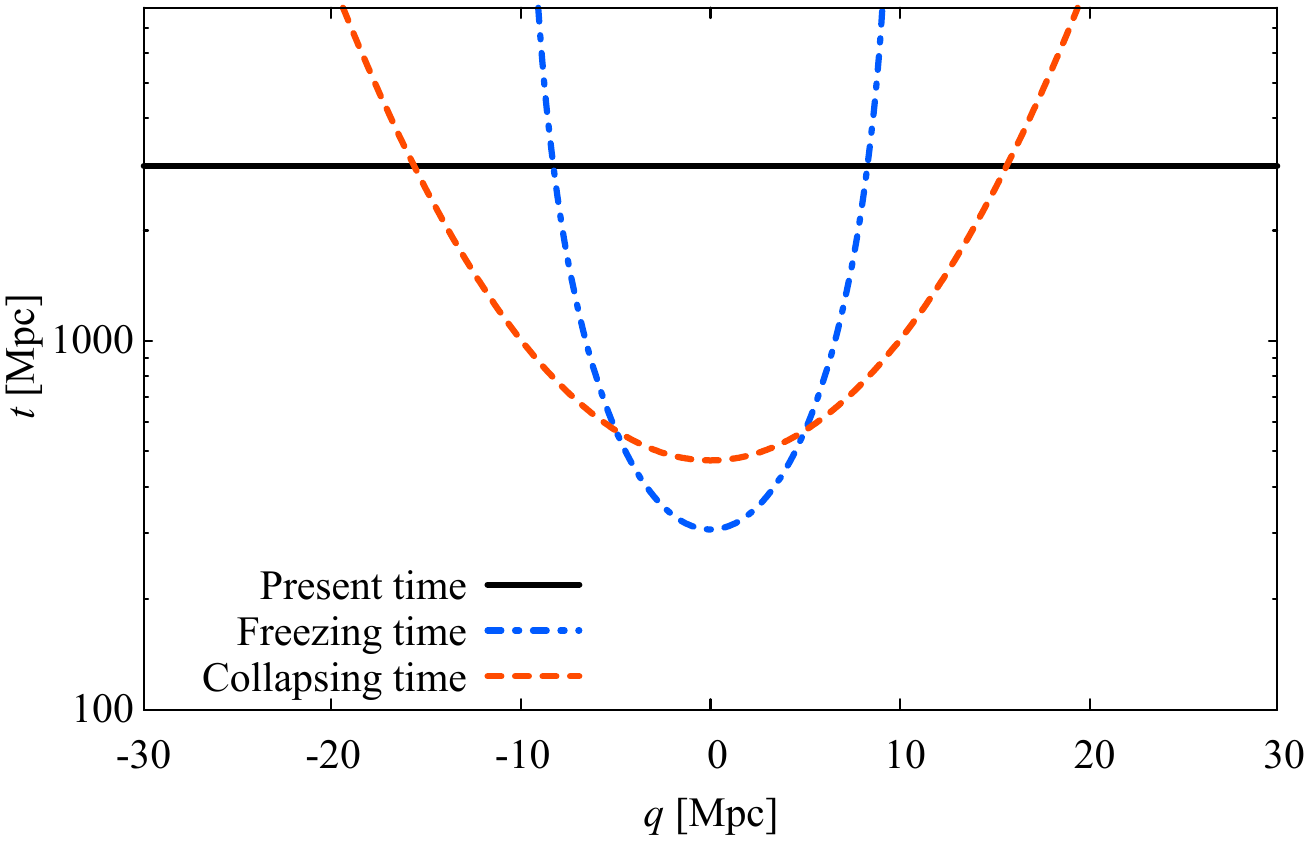} 
        \includegraphics[width=0.49\textwidth]{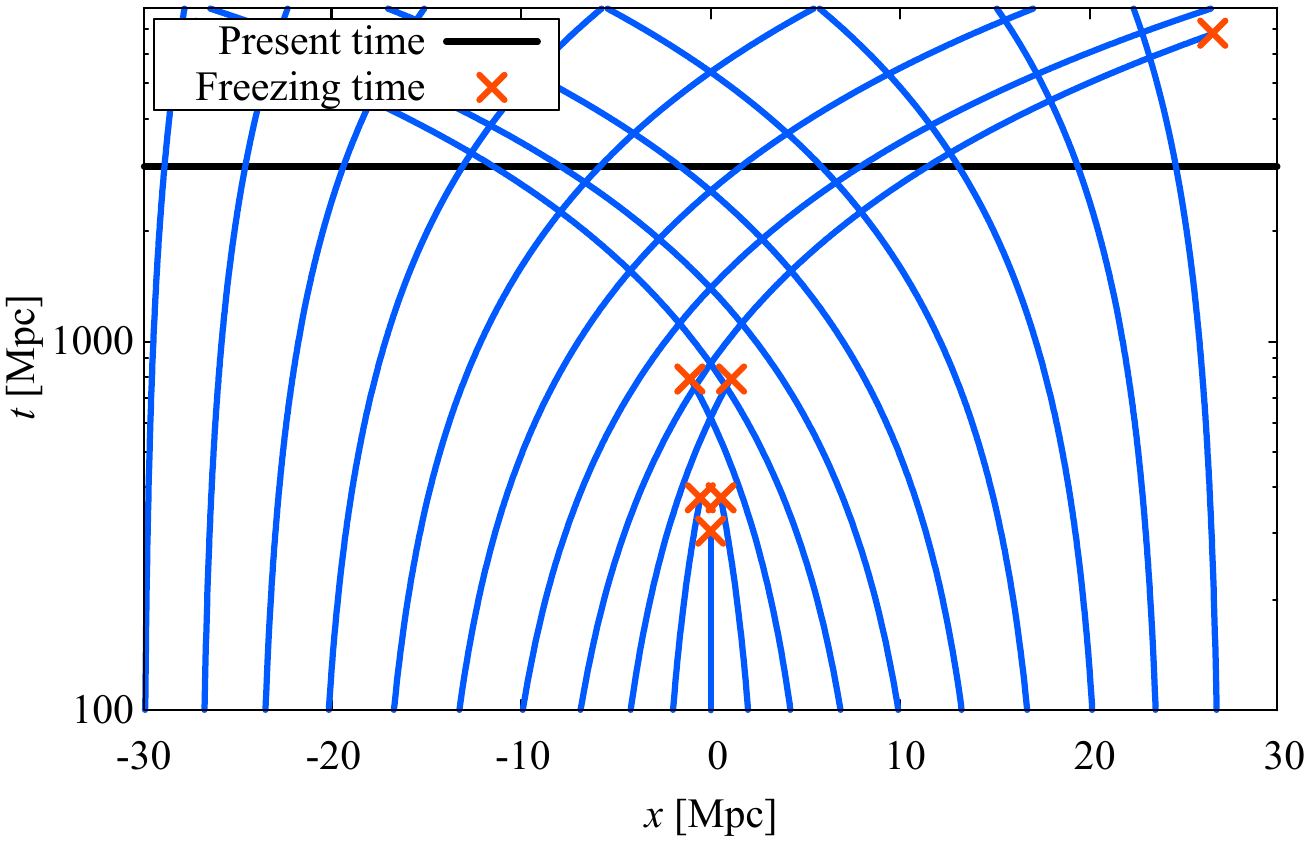} 
        \caption{Left: The collapsing time in the adhesion model and the freezing time in the freezing model for \(A = 10^{-3}\).
        Right: The trajectories of the fluid elements in the freezing model for \(A = 10^{-3}\).
        }\label{fig:timetrajectory}
    \end{figure}
    
Let us explain what is going on in the freezing model. 
In the left panel of Figure~\ref{fig:timetrajectory} the blue dash-dotted curve shows the time when the freezing occurs for each \(q\) in the freezing model, 
while the red dashed curve shows $q_1^*$ and $q_2^*$, which are the boundaries of the collapsed region 
in the adhesion model. 
Here, we set \(A = 10^{-3}\), corresponding to \(|\phi(t=0,q=0)| = 2.6 \times 10^{-5}\).
For the fluid elements of \(|q| \lesssim 5 \,\mathrm{Mpc}\), the freezing time is earlier than the collapsing time, which 
is the time when the fluid element is absorbed by the collapsed region.
On the other hand, for the elements of \(5\,\mathrm{Mpc} \lesssim |q| \lesssim 15 \,\mathrm{Mpc}\), 
the freezing time comes after the collapsing time.
We draw the trajectories of fluid elements in the coordinates $(t,x)$ in the freezing model in the right panel of Figure~\ref{fig:timetrajectory}, 
using Eq.~\eqref{eq:zeldovich}.
The end point of each line denote the freezing time.
In the adhesion model, the collapsing time is the time at which the fluid element reaches \(x = 0\). 
Figure~\ref{fig:timetrajectory} clearly shows that some fluid elements continue to evolve following the Zel'dovich approximation 
even after the caustic formation. Hence, the light ray travels a long distance in the region that is not described well by the Zel'dovich approximation.
In addition, light ray travels on an incorrect path in the freezing model.
Although it uses the Lagrange coordinates \(q\), its coordinates are completely invalid in the frozen collapsed region. 
As can be seen from the right panel of Figure~\ref{fig:timetrajectory}, light ray propagating in the Lagrange coordinate space is seen as passing through the same physical region in Euler coordinates three times,
which is clearly incorrect.

Both models simplify collapsed regions, however, the freezing model has other critical problem, in contrast to the adhesion model. 
To describe collapsed region, the freezing model locally fixes the physical quantities of fluid element, independently of the other fluid elements.
If collapsed region is replaced by some models like the freezing model, accretion flow causes a gap between the velocities of the inside and the outside of the collapsed region.
The method handling fluid element locally does not include the gap properly and leads to a wrong result.
In the adhesion model, the gap is taken into account and consequently the difference between \(H_0^{\mathrm{(B)}}\) and \(H_{0}^{\mathrm{(I)}}\) is small.
These are the reason for the overestimation of the Hubble constant in the freezing model.

\section{Estimation of the Hubble constant by averaging approach}\label{sec:averaging}
To investigate the influence of inhomogeneities of the universe, averaging methods are also used.
If the method is not appropriate, it can lead to apparent effects suggesting a large deviation from the background FLRW spacetime. 
If such a deviation is seen within the present adhesion model, 
it cannot be physical. 
It simply means that the quantities defined by the averaging are not 
directly corresponding to the observables.

Here, we present a simple example.
In the synchronous and comoving gauge, the local expansion rate 
in the plane-symmetric model is expressed as
\begin{align}
    \Theta = 3 \frac{a^{\prime}}{a} - \frac{b^{\prime}\partial_q^2 \Phi}{1 - b\partial_q^2 \Phi}.
\end{align}
Taking \(\Phi \) as a random Gaussian variable 
with the mean \(\langle \Phi \rangle = 0\), 
we can naively obtain the average of the local expansion rate in a region
\begin{align}
    \label{eq:hubble-av1}
     \langle \Theta \rangle  = 3 \frac{a^{\prime}}{a} - b^{\prime}b \langle {(\partial_q^2 \Phi)}^2 \rangle\,,
\end{align}
truncating at the second order in $\Phi$.
The second term in the above expression is proportional to \(b^{\prime} b \). 
Since \(b^{\prime} b \propto T^{1/3}\) in the dust universe, this averaged expansion rate can largely deviate from the background value \(3 {a^{\prime}}/{a}\). The value is always smaller than the background value. 
Actually, the difference between \(\langle \Theta \rangle \) and the background expansion rate is 
\begin{align}
    \frac{\langle \Theta \rangle - 3H}{3H} = - b^{\prime}b \langle {(\partial_q^2 \Phi)}^2 \rangle = \mathcal{O}(10 \epsilon)\, ,
\end{align}
and its magnitude is not negligible.
This large effect does not depend on treatments of the structure formation, for example, even in the adhesion model.
Before caustics occur, the Zel'dovich approximation is the exact solution.
However, in Section~\ref{sec:h0estimation} we have shown that the Hubble constant cannot largely deviate from the background value. 
Hence, the smaller value of this kind of averaged expansion rate should not be interpreted as the backreaction to the Hubble constant. 
This example simply gives us a lesson that we need to directly calculate observable quantities in discussing the effect of inhomogeneities.

\section{Conclusions}\label{sec:conclusion}
We scrutinized the estimates of the Hubble constant using the distance-redshift relation in inhomogeneous universe models.
In this paper, we took into account the inhomogeneities of the universe within the Newtonian cosmology.
We approximated collapsed regions by the adhesion model in which fluid elements obey the Zel'dovich approximation with the infinitesimal viscosity, which is enough to prevent crossing between fluid elements. 
We determined the luminosity distance and the redshift of a source based on the wave vector transported along the null geodesic.
In this way, we demonstrated that the effects of inhomogeneities on the Hubble constant is much smaller than the magnitude of the Hubble tension.

For comparison, we investigated other two ways of the modeling of the inhomogeneous spacetime. 
First, we investigated the freezing model as a modelling of the collapsed region.
In this model fluid elements follow the Zel'dovich approximation when the perturbation is small, and the freezing treatment is adopted after the maximum expansion. In the frozen region some physical quantities are frozen independently from the conditions of the neighboring fluid elements.
In this case, we found that the calculated Hubble constant can significantly deviate from the background value.
The errors, which are already manifest at the Newtonian level, come from the part where the light ray passes through the frozen region, and we concluded that this model is inappropriate to estimate the effects of inhomogeneities on the measured value of the Hubble constant.
The lesson is that use of this kind of modeling can erroneously predict a large correction due to inhomogeneities to the Hubble constant.

Second, we considered a prescription to average the second order perturbation as a method of estimating the deviation of the measured Hubble constant from the homogeneous background value.  
In our example, the difference between the averaged Hubble constant and the background value is much larger than that obtained by considering the light propagation in the adhesion model.
This fact clearly demonstrates that a large correction due to inhomogeneities can be easily erroneously concluded in the estimate of the Hubble constant, if the method of averaging is not really appropriate.

In this paper, we considered a simple setup such that the inhomogeneities is plane-symmetric and the universe is dust dominant.
In this simple setup the Zel'dovich approximation is an exact solution of Newtonian cosmology until caustics occur, and hence we could easily control the errors caused by the treatment of the collapsed region.
Though the model is simple, this setup 
should be enough to get the lesson 
that the over-estimate of the correction due to inhomogeneities to the measured Hubble constant can easily happen.

\section*{Acknowledgements}
This work is supported by JSPS KAKENHI Grant Nos. JP23H00110 and JP20K03928. 

\nocite{*}
\bibliography{bibdata}

\end{document}